\documentstyle[amssymb,prl,aps,twocolumn]{revtex}

\begin{document}
\draft
\title{Distilling multipartite pure states from a finite number of copies of
multipartite mixed states }
\author{Ping-Xing Chen$^{1,2\thanks{%
E-mail: pxchen@nudt.edu.cn}}$and Cheng-Zu Li$^1$}
\address{$^1$ Department of Applied Physics, National University of Defense\\
Technology, Changsha, 410073, P. R. China. \\
$^2$Key Laboratory of Quantum Information, University of Science and\\
Technology of China, Chinese Academy of Sciences, Hefei 230026, P. R. China}
\maketitle

\begin{abstract}
This paper will address the question of the distillation of entanglement
from a finite number of multi-partite mixed states. It is shown that if one
can distill a pure entangled state from n copies of a mixed state $\sigma
_{ABC\cdots }$ there must be at least a subspace in whole Hilbert space of
the all copies such that the projection of $\sigma _{ABC\cdots }^{\otimes n}$
onto the subspace is a pure entangled state. We also show that the
purification of entanglement or distillation of entanglement can be carried
out by local joint projective measurements with the help of classical
communication and local general positive operator valued measurements on a
single particle, in principle. Finally we discuss experimental realizability
of the entanglement purification.
\end{abstract}


A pure entangled state plays perhaps a central role in quantum information.
It not only gives rise to some completely new applications, such as error
correcting code\cite{1}, dense coding\cite{2} and teleportation\cite{3}, but
also has been a source of great theoretical interest, such as non-locality 
\cite{4} and the experimental tests of Bell theorem \cite{5}. However,
interactions with the environment always occur, and a pure entangled state
will became into a mixed state which will degrade the quality of the
entanglement. This raises a problem: how to distill a pure entangled state
from mixed states? Bennett et al \cite{6,7} first proposed an entanglement
purification scheme for a class of two-qubits mixed states. This scheme
needs infinite copies of the mixed state, in principle, and needs {\it %
collective measurements }\cite{8}. Horodecki et al \cite{9} proved that all
mixed entangled states of two qubits can be purified into a singlet if
infinite copies of the mixed state are provided. However, in practice one
cannot get infinite copies of a mixed states. Can one get a pure entangled
state from finite copies of a mixed state? Linden et al \cite{8} showed that
one cannot get a pure entangled state from an individual Werner state by
local operations and classical communication (LOCC). Hent \cite{10}
generalized this result and showed that no scheme can produce a maximally
entangled state of a bipartite system from a generic mixed state of the
system. Recently, we considered the condition for distillability from finite
copies of a mixed state of arbitrary bipartite system \cite{11}, and showed
that one can get a pure entangled state from n copies of a mixed state if
and only if there exists at least a subspace the projection of n copies onto
this subspace is a pure entangled state. We defined the subspace as {\it %
distillable subspace} (DSS). DSS is similar to the decoherence-free space 
\cite{d1}.

This paper will consider the distillation of the entanglement from a single
or finite copies of a multi-partite mixed state. We first analysis the
property of operators for the distillation of entanglement or the
purification of entanglement, and show that any local operation can be
regarded as a combination of three classes of operators, i.e., local
projective operators (LPO), local filter operators (LFO) and local unitary
operators (LUO). Then we prove that if one can distill a pure entangled
state from n copies of a mixed state $\sigma _{ABC\cdots }$ there must be at
least a subspace of whole Hilbert space of the all copies such that the
projection of $\sigma _{ABC\cdots }^{\otimes n}$ onto the subspace is a pure
entangled state. We also show that any operation for the entanglement
purification or the entanglement distillation acting on a Hilbert space of a
quantum system is, in essence, to first project out a subspace which has a
pure entangled state or a highly entangled mixed states, then change the
entanglement of the state projected out. It is shown that only LPO may
produce a pure entangled state from a or many mixed states; LFO may increase
the entanglement of a entangled state, but never can transfer a mixed states
into a pure state. Finally we discuss the experimental realizability of the
entanglement purification.

Consider a mixed entangled state $\sigma _{ABC\cdots }$ shared by separated
parties Alice, Bob and Charles et al$.$ Now we hope distill a pure entangled
state $\left| \Psi \right\rangle $ from $n$ copies of state $\sigma
_{ABC\cdots }.$ Any protocol for the distillation of entanglement from $n$
copies of a mixed state $\sigma _{ABC\cdots },$ $\sigma _{ABC\cdots
}^{\otimes n},$ can be conceived as successive rounds of measurements and
communication by Alice, Bob and Charles et al. After rounds of measurements
and communication, there are many possible outcomes which correspond to many
measurement operators $\{A_i\otimes B_i\otimes C_i\otimes \cdots \}$ acting
on the Alice, Bob and Charles's et al Hilbert space. Each of these operators
is a product of the positive operators and unitary maps corresponding to
Alice's, Bob's and Charles's measurements and rotations, and represents the
effect of the N measurements and communication. If the outcome $i$ occurs,
the given state $\sigma _{ABC\cdots }^{\otimes n}$ becomes:

\begin{equation}
A_i\otimes B_i\otimes C_i\otimes \cdots \sigma _{ABC\cdots }^{\otimes
n}A_i^{+}\otimes B_i^{+}\otimes C_i^{+}\otimes \cdots  \label{1}
\end{equation}
If one can distill a pure entangled state from $\sigma _{ABC\cdots
}^{\otimes n}$, there must be at least an element $A\otimes B\otimes
C\otimes \cdots \in \{A_i\otimes B_i\otimes C_i\otimes \cdots \}$ such that

\begin{equation}
A\otimes B\otimes C\otimes \cdots \sigma _{ABC\cdots }^{\otimes
n}A^{+}\otimes B^{+}\otimes C^{+}\otimes \cdots \rightarrow \left| \Psi
\right\rangle \left\langle \Psi \right| ,  \label{2}
\end{equation}
where $\left| \Psi \right\rangle $ is a pure entangled state. Operator $%
A,B,C $ can be always expressed as: 
\[
A=a_1\left| \phi _1^{\prime }\right\rangle \left\langle \phi _1\right|
+\cdots +a_{n_a}\left| \phi _{n_a}^{\prime }\right\rangle \left\langle \phi
_{n_a}\right| ; 
\]
\begin{equation}
B=b_1\left| \xi _1^{\prime }\right\rangle \left\langle \xi _1\right| +\cdots
+b_{n_b}\left| \xi _{n_b}^{\prime }\right\rangle \left\langle \xi
_{n_b}\right| ;  \label{3}
\end{equation}
\[
C=c_1\left| \eta _1^{\prime }\right\rangle \left\langle \eta _1\right|
+\cdots +c_{n_c}\left| \eta _{n_c}^{\prime }\right\rangle \left\langle \eta
_{n_c}\right| ; 
\]
where $\left\{ \left| \phi _j^{\prime }\right\rangle ,\text{ }j=1,\cdots
,n_a\right\} ,\left\{ \left| \phi _j\right\rangle ,\text{ }j=1,\cdots
,n_a\right\} $ are Alice's two sets of orthogonal vectors; $\left\{ \left|
\xi _l^{\prime }\right\rangle ,\text{ }l=1,\cdots ,n_b\right\} ,\left\{
\left| \xi _l\right\rangle ,\text{ }l=1,\cdots ,n_b\right\} $ are Bob's two
sets of orthogonal vectors; $\left\{ \left| \eta _p^{\prime }\right\rangle ,%
\text{ }p=1,\cdots ,n_c\right\} ,\left\{ \left| \eta _p\right\rangle ,\text{ 
}p=1,\cdots ,n_c\right\} $ are Charles's two sets of orthogonal vectors. $%
0<a_j\leq 1,$ $j=1,\cdots ,n_a;0<b_l\leq 1,$ $l=1,\cdots ,n_b;0<c_p\leq
1,p=1,\cdots ,n_c.$

The operator $A$ in (\ref{3}) can be carried out by following three
operators: 1) a LPO $P_A=\left| \phi _1\right\rangle \left\langle \phi
_1\right| +\cdots +\left| \phi _{n_a}\right\rangle \left\langle \phi
_{n_a}\right| $ which projects out a subspace spanned by bases $\left| \phi
_j\right\rangle ,$ $j=1,\cdots ,n_a;$ 2) a LFO $F_A=a_1\left| \phi
_1\right\rangle \left\langle \phi _1\right| +\cdots +a_{n_a}\left| \phi
_{n_a}\right\rangle \left\langle \phi _{n_a}\right| $ which changes the
relative weights of the components $\left| \phi _j\right\rangle ,$ $%
j=1,\cdots ,n_a;$ 3) a LUO which transfers the Alice's vectors from $\left\{
\left| \phi _j\right\rangle ,\text{ }j=1,\cdots ,n_a\right\} $ to $\left\{
\left| \phi _j^{\prime }\right\rangle ,\text{ }j=1,\cdots ,n_a\right\} ,$
and similarly for $B$ and $C.$ This is the character of the operators for
the distillation of entanglement we shall use below.

{\bf Definitions: }If a pure state $\left| \Psi \right\rangle _{ABC\cdots }$
has reduced density matrices $\sigma _A,\sigma _B,\sigma _C,\cdots $ with
ranks $n_A,n_B,n_C,\cdots ,$ respectively, we say $\left| \Psi \right\rangle
_{ABC\cdots }$ is an $n_A\otimes n_B\otimes n_C\otimes \cdots $ state, where 
$\sigma _A=tr_{BC\cdots }(\left| \Psi \right\rangle _{ABC\cdots
}\left\langle \Psi \right| ),$ and similarly for $\sigma _B,\sigma _C,\cdots
.$ If $\left| \Psi \right\rangle _{ABC\cdots }$ and $\left| \Phi
\right\rangle _{ABC\cdots }$ are of $n_A\otimes n_B\otimes n_C\otimes \cdots 
$ states, we say $\left| \Psi \right\rangle _{ABC\cdots }$ is same dimension
as $\left| \Phi \right\rangle _{ABC\cdots }$

Lemma 1: If two mixed states $\rho _{ABC\cdots }$ and $\rho _{ABC\cdots
}^{\prime }$ in a Hilbert space $H=H_A\otimes H_B\otimes H_C\otimes \cdots $
fulfill 
\begin{equation}
A\otimes B\otimes C\otimes \cdots \rho _{ABC\cdots }A^{+}\otimes
B^{+}\otimes C^{+}\otimes \cdots =\rho _{ABC\cdots }^{\prime },  \label{4}
\end{equation}
where $A,B,C$ are full rank operators to Hilbert spaces $H_A,H_B,H_C,\cdots
, $ respectively, then the rank of $\rho _{ABC\cdots }$ is equal to that of $%
\rho _{ABC\cdots }^{\prime },$ i.e., $r(\rho _{ABC\cdots })=r(\rho
_{ABC\cdots }^{\prime }).$

Proof: We consider a set of eigenstates-decomposition of state $\rho
_{ABC\cdots }$

\begin{eqnarray}
\rho _{ABC\cdots } &=&\sum_{i=1}^{r(\rho _{ABC\cdots })}p_i\left| \Psi
^i\right\rangle _{ABC\cdots }\left\langle \Psi ^i\right| ,  \label{5} \\
p_i &>&0;\sum_ip_i=1.  \nonumber
\end{eqnarray}
Note that 
\begin{equation}
A\otimes B\otimes C\otimes \cdots \left| \Psi ^i\right\rangle _{ABC\cdots
}=\left| \Psi ^{\prime i}\right\rangle _{ABC\cdots },  \label{6}
\end{equation}
then 
\begin{equation}
\rho _{ABC\cdots }^{\prime }=\sum_{i=1}^{r(\rho _{ABC\cdots })}p_i\left|
\Psi ^{\prime i}\right\rangle _{ABC\cdots }\left\langle \Psi ^{\prime
i}\right| .  \label{7}
\end{equation}
Obviously, we have that 
\begin{equation}
r(\rho _{ABC\cdots })\geqslant r(\rho _{ABC\cdots }^{\prime }).  \label{8}
\end{equation}
We introduce operators $A^{\prime },B^{\prime },C^{\prime }$%
\begin{equation}
A^{\prime }=a_1^{\prime }\left| \phi _1\right\rangle \left\langle \phi
_1^{\prime }\right| +\cdots +a_{n_a}^{\prime }\left| \phi
_{n_a}\right\rangle \left\langle \phi _{n_a}^{\prime }\right|  \label{9}
\end{equation}
such that $a_1^{\prime }a_1=\cdots =a_{n_a}^{\prime }a_{n_a},$ and similarly
for $B^{\prime },C^{\prime }.$ Since operators $A,B,C$ and $A^{\prime
},B^{\prime },C^{\prime }$ are full rank, we have that 
\begin{equation}
A^{\prime }\otimes B^{\prime }\otimes C^{\prime }\otimes \cdots \left| \Psi
^{\prime i}\right\rangle _{ABC\cdots }\propto \left| \Psi ^i\right\rangle
_{ABC\cdots }.
\end{equation}
Thus there exists an operator $A^{\prime }\otimes B^{\prime }\otimes
C^{\prime }\otimes \cdots $ such that

\begin{eqnarray}
&&A^{\prime }\otimes B^{\prime }\otimes C^{\prime }\otimes \cdots \rho
_{ABC\cdots }^{\prime }A^{\prime +}\otimes B^{^{\prime }+}\otimes
C^{^{\prime }+}\otimes \cdots  \nonumber \\
&=&\sum_{i=1}^{r(\rho _{ABC\cdots })}p_i^{\prime }\left| \Psi
^i\right\rangle _{ABC\cdots }\left\langle \Psi ^i\right| ,\qquad p_i^{\prime
}>0;\text{ }\sum_ip_i^{\prime }=1,
\end{eqnarray}
this follows that 
\begin{equation}
r(\rho _{ABC\cdots })\leq r(\rho _{ABC\cdots }^{\prime }).  \label{10}
\end{equation}
Combinning (\ref{8}) and (\ref{10}) we finish the proof of Lemma 1.

If one hope distill a pure state with Schmidt numbers n from a bipartite
mixed state $\sigma \in C^n\otimes C^n,$ the operator for this distillation
should be full rank, obviously. From Lemma 1 we can follow that this kind of
distillation is impossible as shown in Ref \cite{10}

Lemma 2: If two pure states $\left| \Psi \right\rangle _{ABC\cdots }$ and $%
\left| \Phi \right\rangle _{ABC\cdots }$ in a Hilbert space $H=H_A\otimes
H_B\otimes H_C\otimes \cdots $ fulfill 
\begin{equation}
A\otimes B\otimes C\otimes \cdots \left| \Psi \right\rangle _{ABC\cdots
}=\left| \Phi \right\rangle _{ABC\cdots },  \label{11}
\end{equation}
where $A,B,C$ are full rank operators to Hilbert spaces $H_A,H_B,H_C,\cdots
, $ respectively, then the ranks of the corresponding reduced density
matrices satisfy $r(\rho _A^\Psi )=r(\rho _A^\Phi ),r(\rho _B^\Psi )=r(\rho
_B^\Phi )$ and $r(\rho _C^\Psi )=r(\rho _C^\Phi ).$

Proof: As shown in appendix A in Ref. \cite{12}, if (\ref{11}) holds, then $%
r(\rho _i^\Psi )\geqslant r(\rho _i^\Phi ),i=A,B,C.$ By the similar proof as
that of Lemma 1, we can introduce operators $A^{\prime },B^{\prime
},C^{\prime }$ in (\ref{9}). Since operators $A,B,C$ and $A^{\prime
},B^{\prime },C^{\prime }$ are full rank, we have that 
\begin{equation}
A^{\prime }\otimes B^{\prime }\otimes C^{\prime }\otimes \cdots \left| \Phi
\right\rangle _{ABC\cdots }\propto \left| \Psi \right\rangle _{ABC\cdots }.
\end{equation}
Thus $r(\rho _i^\Psi )\leq r(\rho _i^\Phi ),i=A,B,C,$ and then $r(\rho
_i^\Psi )=r(\rho _i^\Phi ),i=A,B,C.$ This ends the proof.

Theorem 1:\ Alice, Bob and Charles et al can distill an $n_A\otimes
n_B\otimes n_C\otimes \cdots $pure state $\left| \Psi \right\rangle
_{ABC\cdots }$ from $n$ copies of a mixed state $\sigma _{ABC\cdots }$ if
and only if there is at least one subspace $H_A^{\prime }\otimes H_B^{\prime
}\otimes H_C^{\prime }\otimes \cdots $, $H_A^{\prime }\in H_A^{\otimes n},$ $%
H_B^{\prime }\in H_B^{\otimes n},H_C^{\prime }\in H_C^{\otimes n},$ such
that the projection of state $\sigma _{ABC\cdots }^{\otimes n}$ on this
subspace is an $n_A\otimes n_B\otimes n_C\otimes \cdots $ pure state $\left|
\Psi ^{\prime }\right\rangle _{ABC\cdots }$ or $\left| \Psi \right\rangle
_{ABC\cdots },$ and $\left| \Psi ^{\prime }\right\rangle _{ABC\cdots }$ can
be transferred into the state $\left| \Psi \right\rangle _{ABC\cdots }$ by
LFO and LUO with nonzero probability.

Proof: If there is a subspace $H_A^{\prime }\otimes H_B^{\prime }\otimes
H_C^{\prime }\otimes \cdots $ satisfying the condition in theorem 1 above,
Alice, Bob and Charles can use projective operators $P_A,P_B$ and $P_C$
which project out subspace $H_A^{\prime },H_B^{\prime },H_C^{\prime },$
respectively, to get an $n_A\otimes n_B\otimes n_C\otimes \cdots $ pure
state $\left| \Psi \right\rangle _{ABC\cdots }$ or $\left| \Psi ^{\prime
}\right\rangle _{ABC\cdots }.$ If they get the state $\left| \Psi ^{\prime
}\right\rangle _{ABC\cdots },$ they can transferred the state $\left| \Psi
^{\prime }\right\rangle _{ABC\cdots }$ into the state $\left| \Psi
\right\rangle _{ABC\cdots }$ with nonzero probability by LFO and LUO. Let's
then move to prove the converse. If Alice, Bob and Charles et al can distill
an $n_A\otimes n_B\otimes n_C\otimes \cdots $ pure state $\left| \Psi
\right\rangle _{ABC\cdots }$ from $n$ copies of a mixed state $\sigma
_{ABC\cdots }$, there must be an operator $A\otimes B\otimes C\otimes \cdots 
$ such that 
\begin{eqnarray}
&&A\otimes B\otimes C\otimes \cdots \sigma _{ABC\cdots }^{\otimes
n}A^{+}\otimes B^{+}\otimes C^{+}\otimes \cdots  \nonumber \\
&\rightarrow &\left| \Psi \right\rangle _{ABC\cdots }\left\langle \Psi
\right| .  \label{12}
\end{eqnarray}
Note that projective operators $P_A,P_B,P_C,\cdots $

\[
P_A=\left| \phi _1\right\rangle \left\langle \phi _1\right| +\cdots +\left|
\phi _{n_a}\right\rangle \left\langle \phi _{n_a}\right| ; 
\]
\begin{equation}
P_B=\left| \xi _1\right\rangle \left\langle \xi _1\right| +\cdots +\left|
\xi _{n_b}\right\rangle \left\langle \xi _{n_b}\right| ;
\end{equation}
\[
P_C=\left| \eta _1\right\rangle \left\langle \eta _1\right| +\cdots +\left|
\eta _{n_c}\right\rangle \left\langle \eta _{n_c}\right| , 
\]
act in the subspaces $H_A^{\prime },H_B^{\prime }\ $and $H_C^{\prime }$
spanned by bases $\left\{ \left| \phi _1\right\rangle ,\cdots ,\left| \phi
_{n_a}\right\rangle \right\} _A,\left\{ \left| \xi _1\right\rangle ,\cdots
,\left| \xi _{n_b}\right\rangle \right\} _B$ and $\left\{ \left| \eta
_1\right\rangle ,\cdots ,\left| \eta _{n_c}\right\rangle \right\} _C$ ,
respectively. Then, (\ref{12}) can became 
\begin{eqnarray}
&&A\otimes B\otimes C\otimes \cdots (P_A\otimes P_B\otimes P_C\otimes \cdots
\nonumber \\
&&\sigma _{ABC\cdots }^{\otimes n}P_A\otimes P_B\otimes P_C\otimes \cdots
)A^{+}\otimes B^{+}\otimes C^{+}\otimes \cdots  \nonumber \\
&\rightarrow &\left| \Psi \right\rangle _{ABC\cdots }\left\langle \Psi
\right| .
\end{eqnarray}
$P_A\otimes P_B\otimes P_C\otimes \cdots \sigma _{ABC\cdots }^{\otimes
n}P_A\otimes P_B\otimes P_C\otimes \cdots $ noted as $\sigma _{ABC\cdots
}^{\prime \otimes n}$ is a projection of state $\sigma _{ABC\cdots
}^{\otimes n}$ onto the subspace $H_A^{\prime }\otimes H_B^{\prime }\otimes
H_C^{\prime }.$ The effect of operator $A\otimes B\otimes C\otimes \cdots $
is just to do local unitary operators and local filter operators on whole
subspace $H_A^{\prime }\otimes H_B^{\prime }\otimes H_C^{\prime }.$ Since
operators $P_A,P_B,P_C$ are full-rank to subspaces $H_A^{\prime
},H_B^{\prime },H_C^{\prime },$ respectively, because of Lemma 1 $P_A\otimes
P_B\otimes P_C$ cannot convert a mixed state in subspace $H_A^{\prime
}\otimes H_B^{\prime }\otimes H_C^{\prime }$ into a pure state. So state $%
\sigma _{ABC\cdots }^{\prime \otimes n}$ is a pure state. Because of Lemma 2
and $A,B,C$ being full rank to subspace $H_A^{\prime }\otimes H_B^{\prime
}\otimes H_C^{\prime },$ $\sigma _{ABC\cdots }^{\prime \otimes n}$ is an $%
n_A\otimes n_B\otimes n_C\otimes \cdots $ pure state. From (\ref{12}) we can
follow that $\sigma _{ABC\cdots }^{\prime \otimes n}$ is the state $\left|
\Psi \right\rangle _{ABC\cdots }\left\langle \Psi \right| $ or can be
converted into $\left| \Psi \right\rangle _{ABC\cdots }\left\langle \Psi
\right| $ by LFO and LUO. This ends the proof.

Note that for two pure states $\left| \Psi \right\rangle $ and $\left| \Phi
\right\rangle $ in bipartite system, if they have same Schmidt numbers they
can be transferred into each other by local filter operators. But for
multi-partite system, two $n_A\otimes n_B\otimes n_C\otimes \cdots $ pure
states may belong to two inequivalent classes, such as three qubits's GHZ
state \cite{121} and W state \cite{12}. 
\[
\left| GHZ\right\rangle =\frac 1{\sqrt{2}}(\left| 000\right\rangle +\left|
111\right\rangle ); 
\]
\begin{equation}
\left| W\right\rangle =\frac 1{\sqrt{3}}(\left| 100\right\rangle +\left|
010\right\rangle +\left| 011\right\rangle )
\end{equation}

Theorem 1 above shows that if one can distill a pure entangled state from a
single mixed state $\sigma _{ABC\cdots }$ there must be at least a subspace
such that the projection of the state $\sigma _{ABC\cdots }$ onto the
subspace is a pure entangled state, i.e., there is at least a DSS. To get a
desired pure entangled state, one can first get a pure entangled state by
LPO, then change the pure entangled state into a desired state by LFO and
LUO acting on the subspace projected out by LPO.

Theorem 2: If one can get an $n_A\otimes n_B\otimes n_C\otimes \cdots $ pure
entangled state from the $n$ copies of a mixed state $\sigma _{ABC\cdots }$
, the rank of $\sigma _{ABC\cdots }^{\otimes n}$ is at most $(\dim H_A.\dim
H_B.\dim H_C)^n-n_A.n_B.n_C\cdots +1.$

Proof: If one can get an $n_A\otimes n_B\otimes n_C\otimes \cdots $ pure
entangled state from $n$ copies of a mixed state $\sigma _{ABC\cdots }$
there is at least a DSS $H=H_A^{\prime }\otimes H_B^{\prime }\otimes
H_C^{\prime }$ , $H_i^{\prime }\in H_i^{\otimes n},i=A,B,C,$ and obviously $%
\dim H_i^{\prime }\geqslant n_i,i=A,B,C.$ We always can write down the
matrix $\left[ \sigma _{ABC\cdots }^{\otimes n}\right] $ under the bases
including the full orthogonal bases in the subspace $H^{\perp }$ orthogonal
to the DSS $H,$ and the orthogonal bases in the DSS $H.$ Since the
projection of $\sigma _{ABC\cdots }^{\otimes n}$ onto the DSS $H$ is an $%
n_A\otimes n_B\otimes n_C\otimes \cdots $ pure state $\left| \Psi
\right\rangle $, if we choose a set of orthogonal vectors including $\left|
\Psi \right\rangle $ as the bases of the DSS $H,$ then the matrix $\left[
\sigma _{ABC\cdots }^{\otimes n}\right] $ is at least $\dim H_A^{\prime
}.\dim H_B^{\prime }.\dim H_C^{\prime }-1$ rows zero elements and $\dim
H_A^{\prime }.\dim H_B^{\prime }.\dim H_C^{\prime }-1$ columns zero
elements. Thus the rank of $\sigma _{ABC\cdots }^{\otimes n}$ is at most

\begin{eqnarray}
&&(\dim H_A.\dim H_B.\dim H_C)^n-  \nonumber \\
&&\dim H_A^{\prime }.\dim H_B^{\prime }.\dim H_C^{\prime }+1 \\
&\leq &(\dim H_A.\dim H_B.\dim H_C)^n-n_A.n_B.n_C\cdots +1.  \nonumber
\end{eqnarray}
This ends the proof.

Let's now consider the role of LPO and LFO in the entanglement purification
or the entanglement distillation, respectively. As shown before, any
operator $A\otimes B\otimes C\otimes \cdots $ describing local POV measures
can be regarded as a combination of three classes of operators (namely, LPO,
LFO and LUO). In the case of entanglement distillation if an operator $%
A\otimes B\otimes C\otimes \cdots $ can bring about a pure entangled state $%
\left| \Psi \right\rangle $ from n copies of a mixed state, it is the LPO
that bring about a pure entangled state as same dimension as $\left| \Psi
\right\rangle $ . Because of the Lemma 1, the local filter operator acting
on the whole subspace projected out by LPO can change the entanglement of
the pure state but not can transfer a mixed state into a pure state.
However, if we turn to the problem of whether the entanglement of a mixed
state $\sigma _{ABC\cdots }$ can be increased by LOCC, we find only local
filter operator may increase the entanglement of $\sigma _{ABC\cdots }$ with
non-zero probability. The proof is simple. Let's take the formation of
entanglement as an example. A mixed state $\sigma _{ABC\cdots }$ always can
be expressed as follows \cite{13}:

\begin{equation}
\sigma _{ABC\cdots }=\sum_ip_i\left| \Phi _i\right\rangle \left\langle \Phi
_i\right|  \label{13}
\end{equation}
where $\left| \Phi _i\right\rangle s$ is a set of pure states decomposition
of $\sigma _{ABC\cdots }$ such that $\sum_ip_iE(\left| \Phi _i\right\rangle
) $ is a minimum over all possible decompositions, and thus the formation of
entanglement $E_F(\sigma _{ABC\cdots })$ of $\sigma _{ABC\cdots }$ is $%
\sum_ipE(\left| \Phi _i\right\rangle ).$ $E(\left| \Phi _i\right\rangle )$
is the entanglement of the pure state $\left| \Phi _i\right\rangle $ ( we
now do not know the entanglement of a multi-partite pure state, but here we
can note the entanglement of $\left| \Phi _i\right\rangle $ is $E(\left|
\Phi _i\right\rangle )).$ Any projective operator cannot increase the
entanglement of the pure state $\left| \Phi _i\right\rangle ,$ and cannot
increase the $E_F(\sigma _{ABC\cdots })$ of $\sigma _{ABC\cdots }.$ It is
the LFO that may increase the entanglement of a mixed state. For example, a
mixed state, $\sigma =\lambda [\frac{\sqrt{3}}2\left| 00\right\rangle +\frac %
12\left| 11\right\rangle ]+(1-\lambda )[\left| 01\right\rangle ],$ can be
transferred into a mixed state $\sigma ^{\prime }=\lambda ^{\prime }[\frac{%
\sqrt{2}}2(\left| 00\right\rangle +\left| 11\right\rangle )]+(1-\lambda
^{\prime })[\left| 01\right\rangle ]$ by operator $A=\frac 12\left|
0\right\rangle _A\left\langle 0\right| +\frac{\sqrt{3}}2\left|
1\right\rangle _A\left\langle 1\right| .$ Employing the Wootters's formula 
\cite{13} for calculating the formation of entanglement, we can find that
when $0.9\leq \lambda <1,E_F(\sigma ^{\prime })>E_F(\sigma ).$ Of course, as
claimed in Ref.\cite{8,10} the entanglement of some mixed states, such as
Werner states, cannot be increased by local filter operator.

Let's now look at the role of collective measurements in the entanglement
distillation or the entanglement purification. It is possible that a mixed
state $\sigma $ has no DSS, but $\sigma ^{\otimes n}$ has DSS or $\sigma
^{\otimes n}$ has a subspace the projection of $\sigma ^{\otimes n}$ onto
the subspace is a mixed entangled state with more entanglement than $\sigma $%
. An example of entanglement distillation is a three qubits' mixed state

\begin{eqnarray}
\sigma &=&p[\frac 1{\sqrt{2}}(\left| 0\right\rangle _A\left| 0\right\rangle
_B\left| 0\right\rangle _C+\left| 1\right\rangle _A\left| 1\right\rangle
_B\left| 1\right\rangle _C)]+  \nonumber \\
&&(1-p)[\left| 0\right\rangle _A\left| 1\right\rangle _B\left|
1\right\rangle _C]
\end{eqnarray}
which has no DSS. But $\sigma ^{\otimes 2}$ has a DSS the projection of $%
\sigma ^{\otimes 2}$ onto this DSS is a pure state $\left| \Phi
\right\rangle =$ $\frac 1{\sqrt{2}}(\left| 01\right\rangle _A\left|
01\right\rangle _B\left| 01\right\rangle _C+\left| 10\right\rangle _A\left|
10\right\rangle _B\left| 10\right\rangle _C)$ which can be distilled by
projective operators $P_i=\left| 01\right\rangle _i\left\langle 01\right|
+\left| 10\right\rangle _i\left\langle 10\right| ,i=A,B,C.$ The pure state $%
\left| \Phi \right\rangle $ is shared by six qubits. To make the state $%
\left| \Phi \right\rangle $ be shared by three qubits, we can first use
unitary rotation operations on one of two qubits of A, B, C, respectively.
i.e., $\left| 0\right\rangle _i\rightarrow \frac 1{\sqrt{2}}(\left|
0^{\prime }\right\rangle _i+\left| 1^{\prime }\right\rangle _i),\left|
1\right\rangle _i\rightarrow \frac 1{\sqrt{2}}(\left| 0^{\prime
}\right\rangle _i-\left| 1^{\prime }\right\rangle _i),i=A,B,C,$ then measure
the rotated qubits by local projective measurements. Thus the remained three
qubits of A, B, C share a $\left| GHZ\right\rangle $ state by using a local
phase flip operation. An example of entanglement purification is a Werner
state $\sigma $,

\[
\sigma =F[\left| \Phi ^{+}\right\rangle ]+\frac{1-F}3([\left| \Phi
^{-}\right\rangle ]+[\left| \Psi ^{+}\right\rangle ]+[\left| \Psi
^{-}\right\rangle ]) 
\]
where $\left| \Phi ^{\pm }\right\rangle =\frac 1{\sqrt{2}}(\left|
0\right\rangle _A\left| 0\right\rangle _B\pm \left| 1\right\rangle _A\left|
1\right\rangle _B);\left| \Psi ^{\pm }\right\rangle =\frac 1{\sqrt{2}}%
(\left| 0\right\rangle _A\left| 1\right\rangle _B\pm \left| 1\right\rangle
_A\left| 0\right\rangle _B).$ $\sigma ^{\otimes 2}$ has two subspaces, one
spanned by Alice's bases$\left| 01\right\rangle _A,\left| 10\right\rangle _A$
and Bob's bases $\left| 01\right\rangle _B,\left| 10\right\rangle _B,$
another spanned by Alice's bases$\left| 00\right\rangle _A,\left|
11\right\rangle _A$ and Bob's bases $\left| 00\right\rangle _B,\left|
11\right\rangle _B.$ The projection of $\sigma ^{\otimes 2}$ onto the two
subspaces are a Bell-diagonal state $\sigma ^{\prime }$ with more
entanglement than $\sigma .$ The two examples above means that {\it in the
collection measurements the role of n copies is, in essence, to provide a
bigger Hilbert space where there may exist a DSS or a desired subspace}.

Let's now discuss experimental realizability of the entanglement
purification or the entanglement distillation. The scheme in previous papers 
\cite{6,7} needs very delicate C-NOT operators among different particles.
This is disappointing from an experimental point of view \cite{10}. Our
results show this is not always true. If we can project out the DSS with a
joint projective measurement on particles of two pairs, then the C-NOT is
not necessary, in principle. For example, a new scheme of purification
proposed recently by pan et al \cite{p} used a efficient joint projective
measurements. However, we still cannot say our results is inspiring from the
experimental point of view for the following reasons: 1. the theorem 1 shows
that distilling a pure entangled state requires that a projection of the
original state is pure, namely requires the original state has DSS. This is
a very strong condition which can be satisfied only in the cases where some
especial noise is present \cite{11}; 2. while the multiple copies of a state
which has no DSS may have DSS or desired subspaces, realization of the
multiple copies is very hard although it has been demonstrated for some
systems \cite{p}.

In summary, we discuss the condition for entanglement distillation from
finite copies of multi-partite mixed state, and show that one can distill a
pure entangled state from n copies of a mixed state $\sigma _{ABC\cdots }$
if and only if there must be at least a subspace such that the projection of
the state $\sigma _{ABC\cdots }^{\otimes n}$ onto the subspace is a pure
entangled state. This is a very strong condition and cannot be satisfied in
many cases. If this condition is satisfied and the multiple copies of a
state can be realized the purification or distillation of entanglement can
be carried out by joint projective measurements and local POVM on a single
particle with the help of classical communication, in principle.

\end{document}